\documentclass[final,sort&compress]{aipproc}

\layoutstyle{8x11double}

\begin{document}

\title{Spin Lifetime in Electron-Doped InP Quantum
 Dots}


\author{Y. Masumoto}{
  address={Institute of Physics, University of Tsukuba, Tsukuba 305-8571,
 Japan}}

\author{B. Pal}{
  address={Institute of Physics, University of Tsukuba, Tsukuba 305-8571,
 Japan}
}

\author{S. Oguchi}{
  address={Institute of Physics, University of Tsukuba, Tsukuba 305-8571,
 Japan}
}

\author{M. Ikezawa}{
  address={Institute of Physics, University of Tsukuba, Tsukuba 305-8571,
 Japan}
}

\begin{abstract}
Spin relaxation of electrons doped in InP quantum dots was studied
by means of luminescence pump-probe and Hanle measurements.
Optical pumping makes spins of doped electrons to be oriented in
parallel to the helicity of the circularly polarized excitation.
The luminescence pump-probe showed the spin orientation of the
doped electrons decay on a millisecond time-scale.  Hanle
measurement clarified the spin dephasing relaxation time of doped
electrons is 1.7 ns which is explained by the frozen fluctuation
of nuclear
 spins.
\end{abstract}

\maketitle

Electron spins in quantum dots (QDs) are good candidates for the
quantum memory in the quantum information technology, because
transferring the quantum information from the photon polarization
to the electron spin polarization is direct and one-to-one.
Electrons in QDs are expected to have long spin lifetime and doped
electrons in QDs have infinite lifetime.  We investigated
time-resolved optical orientation of charge tunable InP QDs and
found that the spins of the doped electrons are oriented under
 the
circularly polarized excitation and are preserved in part for up
to sub-millisecond \cite{Ikezawa}.  In this work, we report
nanosecond dephasing time of the spins by means of Hanle
measurement as well as up to millisecond preservation of the spin
by means of photoluminescence (PL) pump-probe in the same QDs.

The samples studied are charge tunable one-layer InP
self-assembled QDs grown on  $n^{+}$-GaAs substrates.  The
excitation source was a cw Ti:sapphire laser and the PL was
detected by a photomultiplier with a 2-channel gated photon
counter (GPC). A photoelastic modulator (PEM) was used to produce right
 and
left circular polarization alternatively at 42 kHz in the
detection path or the excitation path.  The number of electrons in
the QDs was varied with the change of the electric bias applied
across the quantum structure.  Trionic quantum beats in PL showed
that each InP QD contains one doped electron on an average at the
electric bias of $-0.1 \sim -0.2$V \cite{Kozin}.

Under the electric bias of $-0.1$V, the negative circular
polarization (NCP) was observed in PL of InP QDs under the
circularly polarized quasi-resonant excitation and the
longitudinal magnetic field of 0.1T.   Time-resolved study of the
PL circular polarization showed that NCP started at 100ps and is
preserved within the recombination lifetime \cite{Ikezawa}.  The
NCP is explained by considering the optical pumping of the spins
of doped electrons under the circularly polarized excitation and
the simultaneous spin flip-flop process of a photogenerated
electron-hole pair in P-QDs in which the spin of the doped
electron is oriented in parallel to the helicity of the circularly
polarized excitation~\cite{Ikezawa,Pal}.  For the time-resolved
optical orientation of doped electrons in the time region far
beyond the PL lifetime, a pump-probe PL was used
 \cite{Pal}.

\begin{figure}[btp]
\includegraphics[height=.35\textheight]{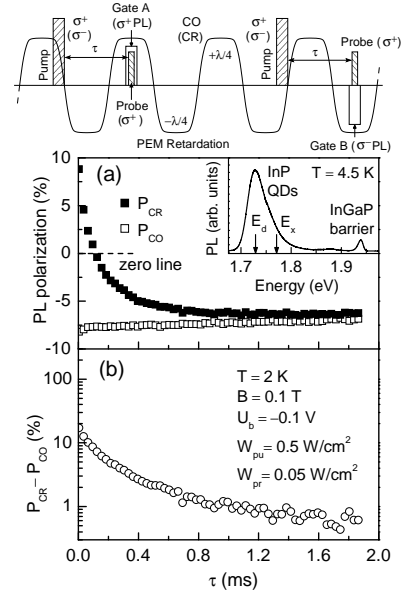}
\caption{ (Upper Sketch): Synchronization of retardation
in a PEM, probe pulses, and
gates of a GPC. (Main Panel): Probe PL
polarization for co- ($P_{\mathrm{CO}}$) and cross-
($P_{\mathrm{CR}}$) circularly polarized pump-probe \textbf{(a)},
and the difference $P_{\mathrm{CR}} - P_{\mathrm{CO}}$ vs. pump-probe delay
 ($\tau$) \textbf{(b)}. PL
spectrum is shown in the
 inset.}
\end{figure}

\begin{figure}[tb]
\includegraphics[height=0.22\textheight]{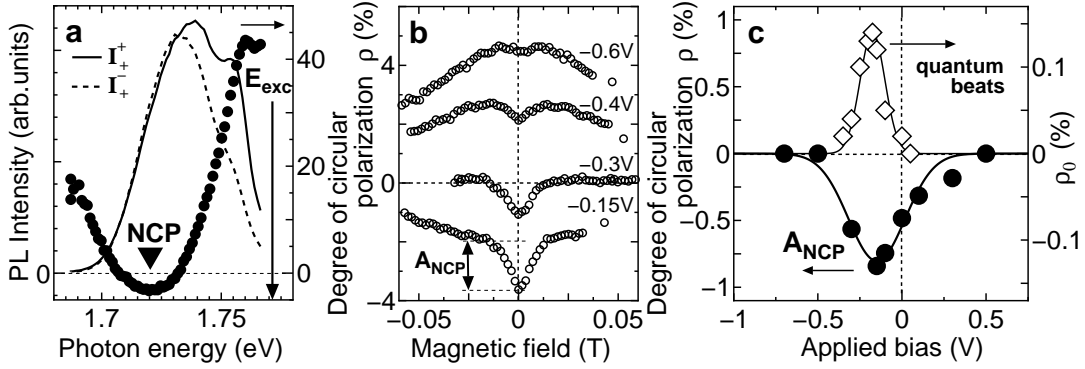}
\caption{\textbf{a}. Circularly polarized PL spectra and
 circular
polarization spectrum of InP QDs.  \textbf{b}. PL circular
polarization of the electrically biased InP QDs under the
transverse magnetic field (Hanle curve).  \textbf{c}. Amplitude of
the sharp Hanle dip and that of trionic quantum beat as a function
of applied electric bias.}
\end{figure}

The timing of the pump, probe pulses and the gate of the photon
counter is shown at the upper part of Fig.1.  We measured the
probe PL polarization for co-circularly polarized
 pump-probe
($P_{\mathrm{CO}}$) and cross-circularly polarized
 pump-probe
($P_{\mathrm{CR}}$) as a function of pump-probe time delay $\tau$.
The experimental data are shown in the main panel of Fig.1.  The
difference $P_{\mathrm{CR}} - P_{\mathrm{CO}}$ is a good measure
of the pump induced spin orientation of the doped electrons.
 A
semi-logarithmic plot of $P_{\mathrm{CR}} - P_{\mathrm{CO}}$ shows
that the spin orientation of doped electrons decays
non-exponentially and that the spin
polarization decays on a millisecond time-scale.  Spin relaxation
rate increases with the increase of the temperature and the
longitudinal magnetic field.  The temperature dependence suggests
two-phonon processes as the dominant spin relaxation mechanism in
QDs at the elevated
 temperatures.

Under quasi-resonant excitation, PL of singly electron doped InP
QDs showed NCP at the Stokes shift of 49 meV even under zero
magnetic field, as shown in Fig.2a.  We measured the Hanle effect
of the sample under the transverse magnetic field.  The Hanle
curves of singly electron doped InP QDs taken in the magnetic
field up to 6 T and under the electric bias of $-0.1$V are
described by the expression, $\rho(B) = A_{0}
 +
A_{\mathrm{NCP}}/[1+(B/B_{1})^2]+A_2/[1+(B/B_{2})^2]+A_3/[1+(B/B_{3})^2]$,
where  $B_{i} = \hbar/g_{i}\mu_{\rm B}T_{2,i}^{\ast} , (i=1,2,3),$
and $T_{2,i}^{\ast}$ is the spin lifetime.  Fitting parameters
are, $A_0=0.10\%$, $A_{\mathrm{NCP}}=-1.43\%$, $B_1=4.5$ mT,
$A_2=-3.97\%$, $B_2$=128 mT, $A_3=2.30\%$, and $B_3$=1.54 T.
Lorentzians having half width at the half maximum (HWHM) of $B_2$
and $B_3$ are observed in neutral InP QDs and are assigned to
depolarization of excitons and holes, respectively~\cite{Comment}.  The
 Hanle
curves precisely measured in the low field regime up to 62.5 mT
are displayed in Figs.2\textbf{b}.  Simultaneously with the
enhancement of the trionic quantum beat, a sharp dip appeared on a
broader Lorentzian in the Hanle curve and the amplitude of the
sharp dip of negative circular polarization, $A_{\mathrm{NCP}}$,
is enhanced around the applied bias of $-0.2$ V, as is shown
 in
Figs.2\textbf{b} and 2\textbf{c}.  The clear coincidence between
NCP and trionic quantum beat shows sharp Hanle dip is reflected by
the long spin dephasing time of doped electrons in QDs.  The
 sharp
Lorentzian dip has HWHM of 4.5 mT.    The HWHM of $B_{1} =$ 4.5 mT
of the sharp Hanle dip corresponds to the electron spin coherence
time of $\tau$ = 1.7 ns, because electron g-factor in InP QDs is
1.5. This observation clearly indicates the spin coherence time of
doped electrons in InP QDs is 1.7 ns. The recombination lifetime
in InP QDs is 250 ps by the time-resolved PL
measurement~\cite{Ikezawa}. The 1.7 ns spin dephasing time of the
doped electron is much longer than the
recombination lifetime.  The time is almost consistent with the
estimated electron spin dephasing time in the randomly distributed
frozen fluctuation of the nuclear hyperfine
field~\cite{Merkulov-hf}.  It is almost consistent with the
dispersion of the hyperfine field caused by nuclear spin
fluctuation measured from a clear zero-field dip (HWHM = 15mT) in
the plot of the circular polarization vs. the longitudinal
magnetic field. Dephasing rate increases with the increase of
temperature, and its temperature dependence suggests the
two-phonon process for dephasing of the electron spin.

This work was supported by the Grant-in-Aid for the Scientific
Research \#13852003 and \#18204028 from the MEXT and "R\&D
promotion scheme funding international joint research" promoted by
NICT of  Japan.


\begin{thebibliography}{9}

\bibitem{Ikezawa}
M.~Ikezawa, B.~Pal, Y.~Masumoto, I.~V.~Ignatiev, S.~V.~Verbin,
 and
I.~Ya.~Gerlovin, \emph{Phys. Rev. B} \textbf{72},
 153302-1--4
(2005).

\bibitem{Kozin}
I.~E.~Kozin, V.~G.~Davydov, I.~V.~Ignatiev, A.~V.~Kavokin,
K.~V.~Kavokin, G.~Malpuech, H.~-W.~Ren, M.~Sugisaki, S.~Sugou,
 and
Y.~Masumoto, \emph{Phys. Rev. B} \textbf{65},
 241312(R)-1--4
(2002).

\bibitem{Pal}
B.~Pal, M.~Ikezawa, Y.~Masumoto, and I.~V.~Ignatiev, \emph{J.
Phys. Soc. Jpn.} \textbf{75}, 54702-1--5
 (2006).

\bibitem{Comment}
Multicomponent Hanle curves for electrons and excitons were observed in
 \emph{Phys. Rev. B} \textbf{66}, 153409-1--4 (2002) by R.~I.~Dzhioev et al.
 and \emph{Phys. Solid State} \textbf{40}, 2024--2030 (1998) by
 V.~P.~Kochereshko et al.

\bibitem{Merkulov-hf}
I.~A.~Merkulov, Al.~L.~Efros, and M.~Rosen, \emph{Phys. Rev. B}
\textbf{65}, 205309-1--8 (2002).


\end{thebibliography}
\end{document}